\documentclass[12pt,a4paper,final]{iopart}
\usepackage[utf8]{inputenc}
\usepackage{iopams}
\expandafter\let\csname equation*\endcsname\relax
\expandafter\let\csname endequation*\endcsname\relax
\usepackage{amsmath}
\usepackage{graphicx,dcolumn,bm,amssymb,amsfonts,xcolor,mathtools,float}
\usepackage{dsfont}  
\usepackage{bbold}   
\usepackage{dutchcal} 
\usepackage[sort&compress,numbers]{natbib}
\usepackage[
pdfstartview=FitBV,
bookmarks=true,
bookmarksopen=true,
colorlinks =true,
linkbordercolor=blue,
linkcolor = blue,
citecolor = blue,
urlcolor=blue]{hyperref}
\usepackage[capitalize]{cleveref} 

\crefname{figure}{Figure}{Figures}

\begin{document}
	
	\title[\small{Mirahmadi et al., Quantum Dynamics of a Planar Rotor Driven by Switched Interactions}]{Quantum dynamics of a planar rotor driven by suddenly switched combined aligning and orienting interactions}
	
	\author{Marjan Mirahmadi,$^1$ Burkhard Schmidt,$^2$ and Bretislav Friedrich$^1$}
	\address{$^1$ Fritz-Haber-Institut der Max-Planck-Gesellschaft, Faradayweg 4-6, D-14195 Berlin, Germany}
	\address{$^2$ Institute for Mathematics, Freie Universit\"{a}t Berlin, Arnimallee 6, D-14195 Berlin, Germany}
	\ead{mirahmadi@fhi-berlin.mpg.de, burkhard.schmidt@fu-berlin.de and bretislav.friedrich@fhi-berlin.mpg.de}
	
	\begin{abstract}
		We investigate, both analytically and numerically, the quantum dynamics of a planar (2D) rigid rotor subject to suddenly switched-on or switched-off concurrent orienting and aligning interactions. We find that the time-evolution of the post-switch populations as well as of the expectation values of orientation and alignment reflects the spectral properties and the eigensurface topology of the planar pendulum eigenproblem established in our earlier work [Frontiers in Physics \textbf{2}, 37 (2014); Eur. Phys. J. D \textbf{71}, 149 (2017)]. This finding opens the possibility to examine the topological properties of the eigensurfaces experimentally as well as provides the means to make use of these properties for controlling the rotor dynamics in the laboratory.
	\end{abstract}

	\section{Introduction}\label{intro}
	Quantum systems driven by external electromagnetic fields have been the subject of countless studies that laid the foundations for our understanding of quantum dynamics and of quantum control \cite{Lemeshko2013,Koch2019}.  Interactions with fields that vary on a time scale much shorter than the characteristic time of a quantum system have been of particular interest as these may lead to analytic results exacted by invoking the sudden approximation \cite{Schiff1968}.
	
	The quantum pendulum, i.e., a rotating dipole, permanent or induced, subject to an external electric field is a prototypical field-driven quantum system whose various incarnations were key to understanding a wide range of problems, ranging from orientation and alignment of molecules by electromagnetic fields \cite{Friedrich1991,Friedrich1991a,Seideman1995,Friedrich1999,Henriksen1999,Cai2001,Stapelfeldt2003,Seideman2005,Daems2005,Kiljunen2005,Leibscher2009,Rouzee2009,Schmidt2014a,Koch2019} to internal rotation and molecular torsion \cite{Herschbach1959,ramakrishna2007,Roncaratti2010,Parker2011,Ashwell2015} to cold atoms in optical lattices \cite{Graham1992,Leibscher2002,Peil2003,Kay2004,Calarco2004,Ayub2009,Kangara2018} to quantum chaos and Anderson localization \cite{Izrailev1980,Fishman1982,Kolovsky1991,Levi2003,Leibscher2004,Tian2010,Floss2013,Floss2014,Bitter2016,Bitter2017}. So far, the majority of experimental and theoretical studies involving the quantum pendulum have been restricted to the purely orienting (proportional to the cosine of the polar angle between the permanent dipole and the external electric field) or aligning interactions (proportional to the cosine squared of the polar angle between the induced dipole and the external electric field), see, e.g., Refs.~\cite{Seideman1995,Henriksen1999,Seideman2001,Cai2001,Leibscher2003,Leibscher2004,Leibscher2004a,Leibscher2009,Shu2010}. Much less attention was paid to the combined orienting and aligning interaction, even though it leads to novel phenomena that cannot be brought forth or explained by either of the two interactions alone, see Refs. \cite{Friedrich1999,Cai2001a,Daems2005,Roncaratti2010,Lemeshko2011,Parker2011,Nielsen2012,Schmidt2014,Schmidt2014a,Schmidt2015,Becker2017,Mirahmadi2018}. 

Herein, we investigate the dynamics of a planar (2D) quantum rotor driven by such combined interactions. 
In particular, we examine combined interactions whose temporal dependence entails a slow (adiabatic) switch-on and a rapid (non-adiabatic) switch-off -- or vice versa -- of the fields that create these interactions \cite{Yan1999,Seideman2001,Kalosha2002,Underwood2003,Stapelfeldt2003,Sussman2006,Muramatsu2009,Chatterley2018}. Such sudden switch-on or switch-off that is effected over a time much shorter than the rotor's rotational period results in a broad rotational wavepacket (for a rapid switch-off) or a wavepacket consisting of a large number of pendular states (for a rapid switch-on). 
	
	By means of both analytic and numerical methods (Split-Operator FFT method via  WavePacket software package \cite{Schmidt2017,Schmidt2018,Schmidt2019}), we were able to carry out the population analysis of the resulting wavepackets as well as of the time-evolution of the observables such as orientation and alignment cosines and the kinetic energy of the planar rotor. Our analysis revealed that these observables mirror the intersections -- avoided and genuine -- of the eigenenergy surfaces spanned by the orienting and aligning parameters whose loci are characterized by an integer termed the topological index. 
	
	Interestingly, the time-independent Schr\"odinger equation  for a rotor subject to the combined interactions belongs to the class of conditionally quasi-exact solvable (C-QES) problems, unlike the purely aligning or orienting interaction. In our previous work, we established that the  condition under which the analytic solutions obtain is intimately connected with the topological index \cite{Schmidt2014,Schmidt2014a,Becker2017,Schatz2018,Mirahmadi2020}. We make use of the analytic solutions to derive aspects of the driven rotor's dynamics analytically.
	
	This paper is organized as follows: In \cref{modeltdse}, we introduce the general Hamiltonian of a linear 2D rotor subject to combined orienting and aligning interactions and discuss its spectral properties. 
	In \cref{switch_on,switch_off}, we examine the quantum dynamics due to rapidly switched-on and switched-off combined interactions, respectively. We evaluate both analytically and numerically the post-switch populations and perform a detailed analysis of the time-evolution of the expectation values of the orientation and alignment cosines. 
In \cref{conclusion}, we provide a summary of the chief results and touch upon possible applications. 
	
\section{Model Hamiltonian}\label{modeltdse}
The Hamiltonian of a planar (2D) rigid rotor with angular momentum $\mathbf{J}=-i\frac{d}{d\theta}$ and moment of inertia $I$ driven by a time-dependent potential $\tilde{V}(\theta,t)$ assumes the general form
	\begin{equation}\label{H_genform}
		\tilde{H}(t) = B \mathbf{J}^2 + \tilde{V}(\theta,t)
	\end{equation}
where $\theta \in [0,\pi]$ is the polar angle and $B = \hbar^2/2I$ the rotational constant. Dividing the time $t$ by $\hbar/B$ results in a dimensionless time $\tau \equiv Bt/\hbar$, whereby the rotational period $t_{\mathrm{r}} =2\pi\hbar/B$ becomes $\tau_{\mathrm{r}}=2\pi$. Moreover, dividing Hamiltonian \eqref{H_genform} through the rotational constant results in a dimensionless Hamiltonian
	\begin{equation}\label{Hdimensionless}
		H(\tau) \equiv \tilde{H}(\tau) / B =\mathbf{J}^2 + V(\theta,\tau)
	\end{equation}
Herein, we consider the dimensionless interaction potential $V(\theta,\tau)\equiv \tilde V(\theta,\tau)/B$ that drives the rotor to be comprised of a combined interaction
	\begin{equation}\label{V_genform}
		V(\theta,\tau) =  -\eta(\tau)\cos\theta-\zeta(\tau)\cos^2\theta
	\end{equation}
with $\eta(\tau)$ and $\zeta(\tau)$ dimensionless parameters characterizing, respectively, the strengths of the orienting, $\cos\theta$, and aligning, $\cos^2\theta$,  interactions. 
The corresponding time-dependent Schr\"{o}dinger equation (TDSE) for the wavefunction $\psi(\theta,\tau)$ then takes the form
	\begin{equation}\label{tdse}
		i \frac{\partial}{\partial\tau} \psi(\theta,\tau) = \left[ -\frac{\mathrm{d}^2}{\mathrm{d}\theta^2} - \eta(\tau) \cos\theta - \zeta(\tau) \cos^{2}\theta\right] \psi(\theta,\tau)
	\end{equation}

The combined orienting and aligning potential \eqref{V_genform} is a $2\pi$-periodic function whose shape can be controlled by tuning the orienting and aligning parameters $\eta(\tau)$ and $\zeta(\tau)$. This shape can be varied from a single-well ($|\eta(\tau)|>2\zeta(\tau)$) to an asymmetric double-well ($|\eta(\tau)|<2\zeta(\tau)$) over the interval  $\theta \in [0, \pi]$. Herein, we only investigate the asymmetric double-well regime that pertains to rotors representing linear molecules \cite{Haertelt2008}; see also \cite{Magnus2004,Roncaratti2010,Mirahmadi2020}. The choice of the sign of $\eta$ is arbitrary since it is equivalent to a shift of the potential by $\pi$ in $\theta$. Hence without a loss of generality, we assume $\eta\le0$, in which case $V(\theta)$ has a local minimum at $\theta = 0$ and a global minimum at $\theta = \pi$. 

The stationary counterpart of TDSE \eqref{tdse} obtains for time-independent or slow-varying interaction parameters $\eta(\tau)$ and $\zeta(\tau)$,
\begin{equation}\label{tise}
		\left[ -\frac{\mathrm{d}^2}{\mathrm{d}\theta^2} - \eta \cos\theta - \zeta \cos^{2}\theta\right] \phi_n(\theta) = \epsilon_n \phi_n(\theta)
	\end{equation}  
with eigenfunctions $\phi_n(\theta)$ and dimensionless eigenenergies $\epsilon_n$ (expressed in terms of the rotational constant $B$). ``Slow'' (or adiabatic) corresponds to $\tau \gg\tau_{\mathrm{r}}$, in which case we write $\eta(\tau)\rightarrow \eta$ and $\zeta(\tau)\rightarrow \zeta$.

For $\eta=\zeta = 0$, the time-independent Sch\"odinger equation (TISE) ~\eqref{tise} becomes that of a free planar rotor (or, equivalently, a particle on a ring) with eigenvalues $\epsilon_J=J^2$ and eigenfunctions  $\mathrm{e}^{iJ\theta}/\sqrt{2\pi}$, where the angular momentum quantum number, $J$, is an integer, $ J\in\{0,\pm 1,\pm 2,\cdots\}$, for a periodic boundary condition on the interval $\theta \in[0,2\pi]$. The free-rotor states have a definite parity, given by $(-1)^J$, with the parity transformation defined as $\theta \mapsto -\theta$. 
	
In what follows, we investigate a particular time dependence of the combined orienting and aligning interaction \eqref{V_genform}, namely a sudden switch-off or switch-on, where ``sudden" corresponds to $\tau \ll \tau_{\mathrm{r}}$. In our treatment, we make use of the stationary solutions of TISE \eqref{tise} for constant values of the orienting and aligning parameters $\eta$ and $\zeta$.  The corresponding eigenproblem has been investigated in detail numerically \cite{Schmidt2014} as well as via supersymmetry and $sl(2)$-algebraic approach \cite{Schmidt2014a,Becker2017,Mirahmadi2020}. In the next two subsections, we  summarize the main results.

\subsection{Symmetry properties of the eigenstates}\label{labeling}

The TISE~\eqref{tise} can be mapped onto the Whittaker-Hill differential equation, which is a special case of the Hill equation \cite{Magnus2004}. The Whittaker-Hill equation has four sets of linearly independent  solutions, each set corresponding to an irreducible representation of the planar pendulum's $C_{2v}$ point group~\cite{Becker2017}. Two sets (labelled $A_1$ and $A_2$) satisfy the periodic boundary condition on the interval $\theta \in[0,2\pi]$ and two sets (labelled $B_1$ and $B_2$) satisfy the antiperiodic boundary condition on the said interval. Henceforth, we only consider the $A_1$ and $A_2$ sets of solutions that satisfy the periodic boundary condition \cite{Becker2017,Mirahmadi2020},
	\begin{align}\label{eig_A}
        \phi_{n}^{(A_1)}(\theta) =& \left(N^{(A_1)}_{n}\right)^{-1/2} e^{\sqrt{\zeta}\cos\theta}                    
        \sum_{\ell=0}^{\infty}  v_{n,\ell} \cos^{2\ell}  \frac{\theta}{2} ~, \nonumber \\
        \phi_{n}^{(A_2)}(\theta) =&  \left(N^{(A_2)}_{n}\right)^{-1/2}
        e^{\sqrt{\zeta}\cos\theta} \sin\theta  \sum_{\ell=0}^{\infty} \tilde{v}_{n,\ell} \cos^{2\ell}\frac{\theta}{2} 
    \end{align}
Their normalization factors $N^{(A_1)}_{n}$ and $N^{(A_2)}_{n}$ are given by \cref{N_A1A2} of \ref{pendular_int}.
	The constants $v_{n,\ell}$ and $\tilde{v}_{n,\ell}$ are components of the eigenvectors of the irreducible matrix representations associated with the $A_1$ or $A_2$ symmetries (further details are given in Ref.~\cite{Becker2017}). 
	
	Note that the integer $n$ in TISE~\eqref{tise} labels a state irrespective of its symmetry.  For the sake of notational simplicity, we drop the label $\Gamma \in \{A_1,A_2\}$ unless a state's symmetry needs to be emphasized.
	
	\begin{figure}
		\centering
		\includegraphics[scale=0.55]{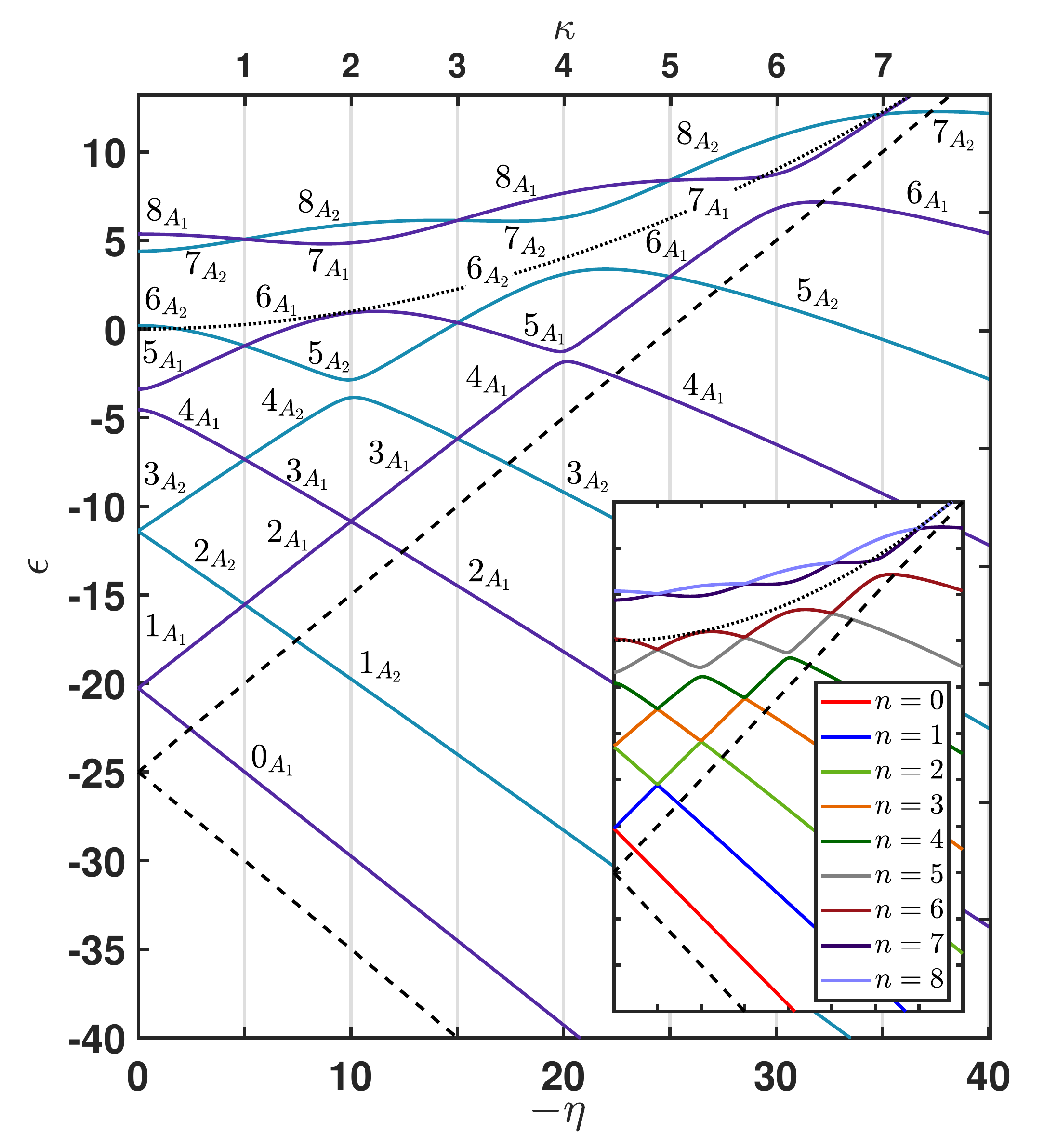} 
		\caption{\label{spec}Eigenenergies of the nine lowest pendular states with symmetries $A_1$ and $A_2$ for $\zeta = 25$. The combined orienting and aligning potential $V(\theta)$ is an asymmetric double-well with a local minimum (upper black dashed line) and a global minimum (lower black dashed line), and a maximum (black dotted curve). The upper abscissa shows the topological index $\kappa$. The inset shows the labelling of the states as used in \eqref{tise}. The eigenenergies shown have been calculated using the WavePacket software package \cite{Schmidt2017,Schmidt2018,Schmidt2019}.}
	\end{figure}

\subsection{Eigenenergies and their topology}\label{eigenenergies}
\cref{spec} shows the dependence of the eigenvalues $\epsilon_n$ on the orienting interaction parameter $\eta$ for a fixed value of the aligning parameter $\zeta = 25$. All states bound by the double-well potential ($|\eta(\tau)|<2\zeta(\tau)$) occur as doublets split by tunnelling through the potential's equatorial barrier, located at $\theta = \arccos[-\eta/(2\zeta)]$, and each doublet has a particular symmetry, either $A_1$ or $A_2$. At $\eta=0$, the double-well potential is symmetric and the tunnelling splitting is solely due to the aligning interaction. As can be seen in \cref{spec}, the splitting becomes ever more apparent as $n$ increases, see the deepest-bound doublet comprised of the $0_{A_1}$ \& $1_{A_1}$ states and the barely-bound doublet $4_{A_1}$ \& $5_{A_1}$. The parity of the states created by the aligning interaction alone is $(-1)^n$, i.e., the same as that of the free-rotor states with which they adiabatically correlate, see Ref. \cite{Friedrich1991a}. At $|\eta|>0$, the double well-potential becomes asymmetric and the tunnelling splitting increases linearly with increasing $|\eta|$, whereby the eigenenergy of the lower member of a given tunnelling doublet decreases (high-field seeker) and that of the upper member increases (low-field seeker). As a result, the eigenenergy of the upper member of a lower tunnelling doublet is bound to intersect the eigenenergy of the lower member of a higher tunnelling doublet. However, only eigenenergies of states of different symmetry, i.e., $A_1$ and $A_2$, may undergo genuine intersections, i.e., become degenerate at the intersection point. This is consistent with the coexistence theorem for the Whittaker-Hill equation \cite{Magnus2004}. In contrast, the eigenenergies of states of the same symmetry, i.e., either $A_1$ and $A_1$ or $A_2$ and $A_2$, may only exhibit avoided crossings \cite{Neumann_Wigner_1929}.  The resulting pattern of eigenenergies as a function of $\eta$ at a fixed $\zeta$ is shown in the main panel of \cref{spec}. We note that the parity of the pendular states created either by the orienting interaction alone or by the combined orienting and aligning interaction is indefinite (mixed). This is indeed the reason why these states can exhibit orientation (i.e., have a non-vanishing orientation  cosine). 

Remarkably, the loci of the intersections, whether genuine or avoided, have a simple analytic form
\begin{align}\label{kappa}
|\eta| = \kappa \sqrt{\zeta}
\end{align} 
where $\kappa =1,2,3,\cdots$ is termed the topological index \cite{Schmidt2014}. Its values are shown on the upper abscissa of \cref{spec}. Odd values of the topological index pertain to genuine crossings and even ones to avoided crossings.  Furthermore, the state label $n$, see also the inset of \cref{spec}, gives the number of intersections that a state with that label partakes in. 

As shown in our previous work \cite{Schmidt2014a,Becker2017}, the integer values of the topological index represent a condition under which the planar pendulum eigenproblem becomes analytically solvable (C-QES) and, at the same time, provides the number of analytic solutions. The pendular eigenfunctions given by Eq. \eqref{eig_A} with the upper limit of the summation determined by $\kappa$ are analytic.

The change of the symmetry of a state at a genuine intersection results in dramatic changes in the localization of its wavefunction.  However, such changes also arise at the avoided crossings, although the symmetry of the states involved remains the same. This is because the localization of the eigenfunctions around $\theta=0$ and $\theta=\pi$ interchanges at the avoided crossing due to the increase in the global minimum and a decrease in the local minimum of the potential $V(\theta)$. The change in the localization of the wavefunction in the vicinity of an avoided crossing is illustrated in \cref{avoided}.
As shown below, either type of relocalization of the wavefunction at a crossing plays out in the dynamics of the driven rotor.
	
	\begin{figure}
		\centering
		\includegraphics[scale=0.55]{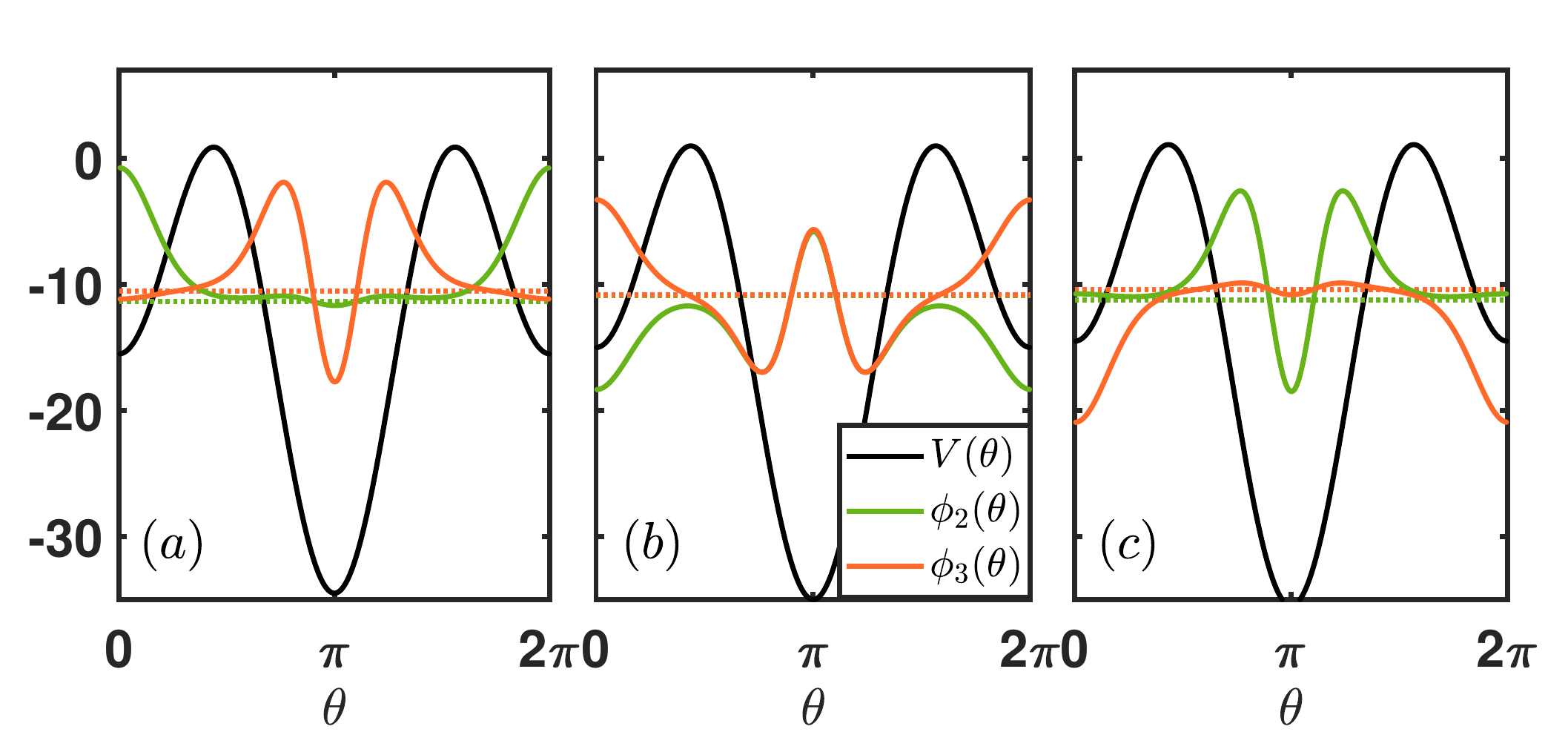} 
		\caption{\label{avoided}Wavefunctions of the $\phi_2$ (green curve) and $\phi_3$ (orange curve) eigenstates of the combined pendular potential $V(\theta)$ (black curve) for $\zeta =25$ and $(a)~ \eta = -9.5 \equiv\kappa=1.9$, $(b)~ \eta = -10 \equiv\kappa=2$, and $(c)~ \eta = -10.5 \equiv\kappa=2.1$. Reprinted, with permission, from Ref.~\cite{Mirahmadi2020}.}
	\end{figure} 

    We note that the pointed gaps and the concomitant abrupt changes in the localization of the wavefunctions at the intersections as illustrated in \cref{avoided} are characteristic of energy levels well-below the maximum of the potential (but still above the local minimum). For barely-bound levels (close to the potential's maximum), the gaps of the eigenenergies at the crossings become rather blunt and their wavefuntions vary more gradually.
    	
	Finally, we emphasize what may have been obvious all along, namely that for a purely orienting or purely aligning interaction, avoided or genuine crossings do not arise.	
\section{Results and discussion}\label{results}
	\subsection{Dynamics driven by a rapidly switched-off combined interaction}\label{switch_off}
\begin{figure}
		\centering
		\includegraphics[scale=0.5]{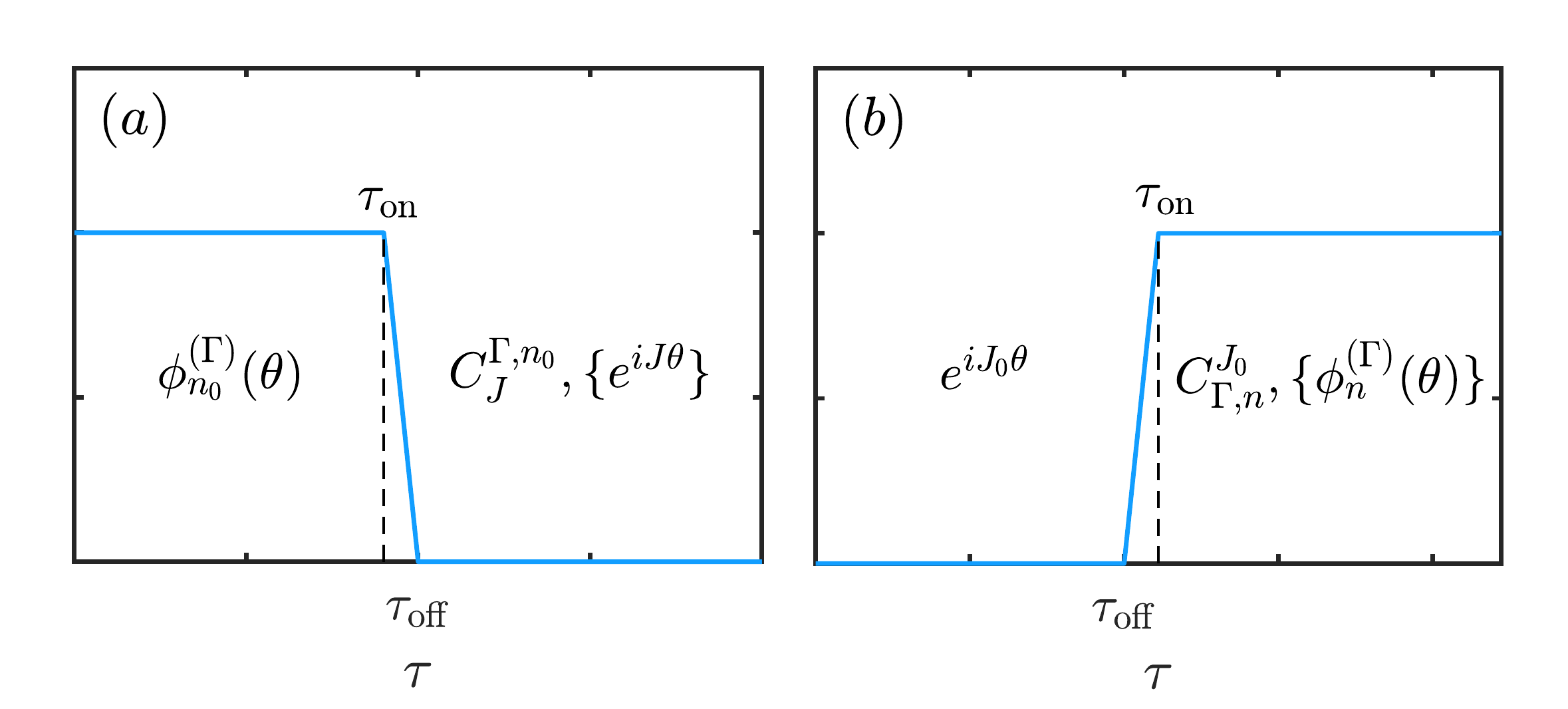}
		\caption{\label{switches}Schematic illustrating the temporal evolution for (a)  a sudden switch-off and (b) a sudden switch-on. Also shown are the pre-switch wavefunctions as well as the post-switch basis functions along with the post-switch expansion coefficients. See text.}
	\end{figure}

	Let us now consider the dynamics of the planar pendulum that arises upon a sudden switch-off of the concurrent orienting and aligning interactions. This case would be brought about by adiabatically switching-on the combined interaction and then switching it off during a time interval $\tau_{\mathrm{on}}-\tau_{\mathrm{off}}\equiv \Delta \tau \ll \tau_{\mathrm{r}}$. Panel (a) of \cref{switches} illustrates this sequence.
By integrating TDSE~\eqref{tdse} over the duration of the switch-off $\Delta \tau$, we obtain
	\begin{equation}\label{tdseint}
		\psi(\theta,\tau_{\mathrm{on}}) -  \psi(\theta,\tau_{\mathrm{off}})  = -i\int_{\tau_{\mathrm{on}}}^{\tau_{\mathrm{off}}}H(\tau) \psi(\theta,\tau)d\tau 
	\end{equation}
	Since the integrand on the right-hand side of \cref{tdseint} is finite, the integral is of the order of the switching duration $\Delta \tau \rightarrow 0$, in which case \cref{tdseint} gives $\psi(\theta,\tau_{\mathrm{on}}) = \psi(\theta,\tau_{\mathrm{off}}) $. Assuming that before the switch-off the system was in a pendular state $n=n_0$, we have $\psi(\theta,\tau_{\mathrm{off}}) = \phi_{n_0}^{(\Gamma)}(\theta)$ with $\Gamma\in\{A_1,A_2\}$.
	Hence the problem reduces to expanding the pendular wavefunction in terms of the free rotor wavefunctions. This expansion yields the post-switch wavefunction
	\begin{equation}\label{wf_expan_off}
		\psi^{(\Gamma)}(\theta,\tau) = \frac{1}{\sqrt{2\pi}} \sum_{J=-\infty}^{+\infty}C^{\Gamma,n_0}_J \mathrm{e}^{-iJ^2\tau} \mathrm{e}^{iJ\theta}
	\end{equation} 
	where
	\begin{align}\label{C_offf}
		C^{\Gamma,n_0}_J = \frac{1}{\sqrt{2\pi}}\int_0^{2\pi}d\theta \mathrm{e}^{-iJ\theta} \phi_{n_0}^{(\Gamma)}(\theta)  
	\end{align}
with $\phi_{n}^{(\Gamma)}$ defined by \cref{eig_A}. We note that  during its field-free propagation, the resulting wavepacket, $\psi(\theta,\tau>\tau_{\mathrm{off}})$, maintains the  symmetry $\Gamma$ of the pre-switch pendular state $\phi_{n_0}^{(\Gamma)}$. Therefore, for initial pendular states $\phi_{n_0}^{(A_1)}$  or $\phi_{n_0}^{(A_2)}$, \cref{wf_expan_off} consists of terms with $C_{J}^{A_1,n_0}$ or $C_{J}^{A_2,n_0}$ coefficients, except at the genuine crossings, where $\psi(\theta,\tau)$ includes a mixed distribution of rotational states with both $C_{J}^{A_1,n_0}$ and $C_{J}^{A_2,n_0}$. 
	
	The analytic expressions for the coefficients $C^{\Gamma,n_0}_J$ along with a summary of the procedure used in Ref.~\cite{Mirahmadi2020} to evaluate them are given in \ref{cal_GJ}, \cref{C_A1,C_A2}. These expressions furnish the rotational wavepackets in closed form (the $C_{J}^{\Gamma,n_0}$ coefficients for $n_0=0$ were found to converge to zero for $J \lesssim 50$). 		
		
	We note that even though \cref{C_A1,C_A2} have been obtained analytically by virtue of the analytic form of the pendular eigenfunctions, Eq. \eqref{eig_A}, i.e., for states with $n<\kappa$ where $\kappa$ is an odd integer (C-QES condition), Eqs. \eqref{wf_expan_off} and \eqref{C_off} are applicable to numerical calculations as well.


\subsubsection{Population analysis}\label{pop_soff}
\begin{figure}
		\centering
		\includegraphics[scale=0.5]{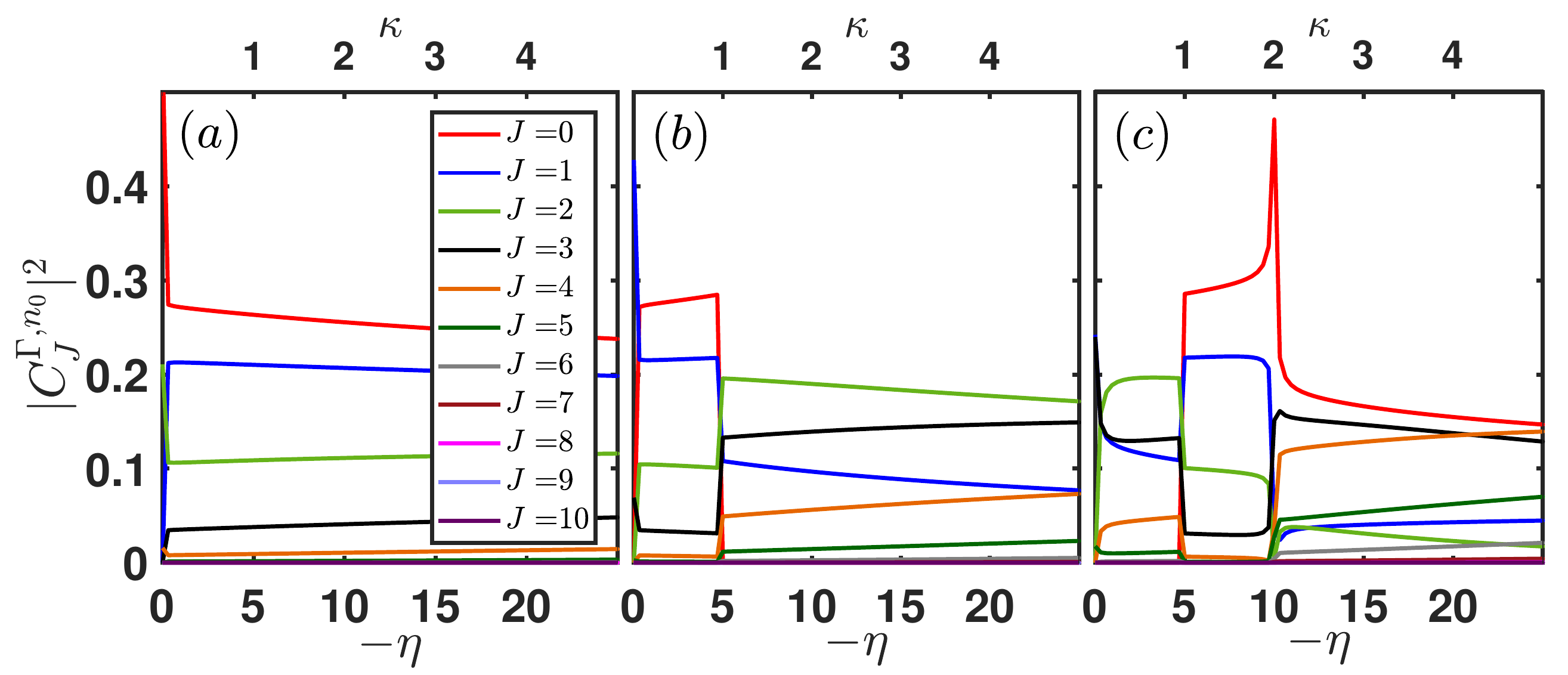}
		\caption{\label{coeff_switch_off}Populations of rotational states, $|C_{J}^{\Gamma,n_0}|^2$, as a function of $\eta$ at a fixed $\zeta=25$ after a sudden switch-off of the combined orienting and aligning interaction. Panels  (a), (b), and (c) correspond to initial pendular states $n_0 = 0,1,2$, respectively. The upper abscissas show the values of the topological index $\kappa$. The color-coding of pendular states $n$ is the same as in \cref{spec}. Adapted, with permission, from Ref.~\cite{Mirahmadi2020}.}
	\end{figure}

	The populations, $|C_{J}^{\Gamma,n_0}|^2$, of the rotational states that make up the wavepacket of Eq. \eqref{wf_expan_off}  for different initial pendular states $\psi(\theta,\tau_{\mathrm{off}}) = \phi_{n_0}^{(\Gamma)}(\theta)$ are shown in \cref{coeff_switch_off} as functions of $\eta$ for $\zeta= 25$.  Only states with $J\ge0$ are displayed as $|C_{J}^{\Gamma,n_0}|^2$ are symmetric with respect to $J=0$.  
    
\cref{C_offf} provides a key to explaining the populations of the free-rotor states produced by the sudden switch-off: the $J$-populations  are given by the overlap of the free rotor and pendular wavefunctions and thereby connected to the symmetries $\Gamma$. Thus in the vicinity of $\eta = -5$ (i.e., of a genuine crossing at $\kappa = 1$), the populations $|C_{J}^{\Gamma,n_0}|^2$ in panels (b) and (c) of \cref{coeff_switch_off} for the initial states $n_0=1$ and $n_0=2$ (cf. \cref{spec}) are seen to exhibit sudden drops and rises due to the corresponding symmetry swaps, $A_1 \leftrightarrow A_2$. Such sudden population changes are absent for the ground state $n_0=0$, see panel (a), as it always pertains to the $A_1$ symmetry. 

In the vicinity of  $\eta=-10$ (i.e., of an avoided crossing at $\kappa=2$), the localization of the pendular states undergoes dramatic changes due to the variation of the potential that affects the overlaps with the rotational wavefunctions and thus the field-free rotational populations, see panel (c) of \cref{coeff_switch_off}.
However, for states above the maximum of the potential $V(\theta)$ (i.e., for hindered-rotor states instead of pendular states) the wavepackets are no longer localized and the sudden changes of the free-rotor populations are absent \cite{Mirahmadi2020}. 
    
We note that for initial states with symmetry $A_2$, the rotational state $J=0$ is not populated, i.e., the corresponding $|C_{0}^{A_2,n_0}|^2$ vanishes identically, see, e.g., panel (b) for $\eta<-5$.

	\subsubsection{Evolution of the expectation values}\label{so_ori_align}
	Given the populations of the post-switch free-rotor states, as established in the previous subsection, we now examine the temporal behaviour of the rotational kinetic energy, orientation, and alignment subsequent to the sudden turn-off of the combined orienting and aligning interactions. 
	
First we note that the post-switch (dimensionless) kinetic energy of the system
	\begin{equation}\label{kinetic_switchoff}
		\left\langle \mathbf{J}^2 \right\rangle=\sum_{J} J^2 \left|C_{J}^{\Gamma,n_0}\right|^2 
	\end{equation} 
remains constant -- a conclusion we could have anticipated based on energy/angular momentum conservation; however, it also checks out computationally.  We note that by tuning the interaction parameters $\eta$ and $\zeta$ prior to the switch-off, we can preordain the kinetic energy of the system in the post-switch, field-free regime, see \cref{coeff_switch_off} and Ref.~\cite{Mirahmadi2020}). 
	
	In order to examine the post-switch evolution of the orientation and alignment characterized, respectively, by the time-dependent expectation values $\left\langle \cos\theta  \right\rangle(\tau)$ and $\left\langle \cos^2\theta  \right\rangle(\tau)$, we evaluate integrals of the form $\langle \psi(\theta,\tau) | \cos^{\gamma} \theta | \psi(\theta,\tau) \rangle$ with $\gamma \in \{ 1,2 \}$. Introducing $\Delta J \equiv J'-J$ and making use of \cref{wf_expan_off}, the integral for $\gamma =1$ is non-vanishing only if $|\Delta J|=1$, yielding \cite{Mirahmadi2020}
	\begin{align}\label{orien_soff}
		\left\langle \cos\theta  \right\rangle (\tau) = \sum_{J=-\infty}^{+\infty} \frac{1}{2}\left\{\mathrm{e}^{-i(2J+1)\tau}\left(C^{\Gamma,n_0}_{J+1}\right)^* C^{\Gamma,n_0}_J + \mathrm{e}^{i(2J-1)\tau}\left(C^{\Gamma,n_0}_{J-1}\right)^*C^{\Gamma,n_0}_J \right\} 
	\end{align}
	with $\Gamma\in\{A_1,A_2\}$. For $\gamma=2$, the integral vanishes unless $|\Delta J|=0 , 2$, yielding
	\begin{align}\label{align_soff}
		\left\langle \cos^2\theta  \right\rangle (\tau) =& \sum_{J=-\infty}^{+\infty} \frac{1}{4} \left\{ \mathrm{e}^{-4i(J+1)\tau}\left(C^{\Gamma,n_0}_{J+2}\right)^* C^{\Gamma,n_0}_J +  \mathrm{e}^{4i(J-1)\tau} \left(C^{\Gamma,n_0}_{J-2}\right)^* C^{\Gamma,n_0}_J \right\} \nonumber \\
		&+ \frac{1}{2} \sum_{J=-\infty}^{+\infty} \left|C^{\Gamma,n_0}_J\right|^2 
	\end{align}
    
\begin{figure}
    	\centering
    	\includegraphics[scale=0.45]{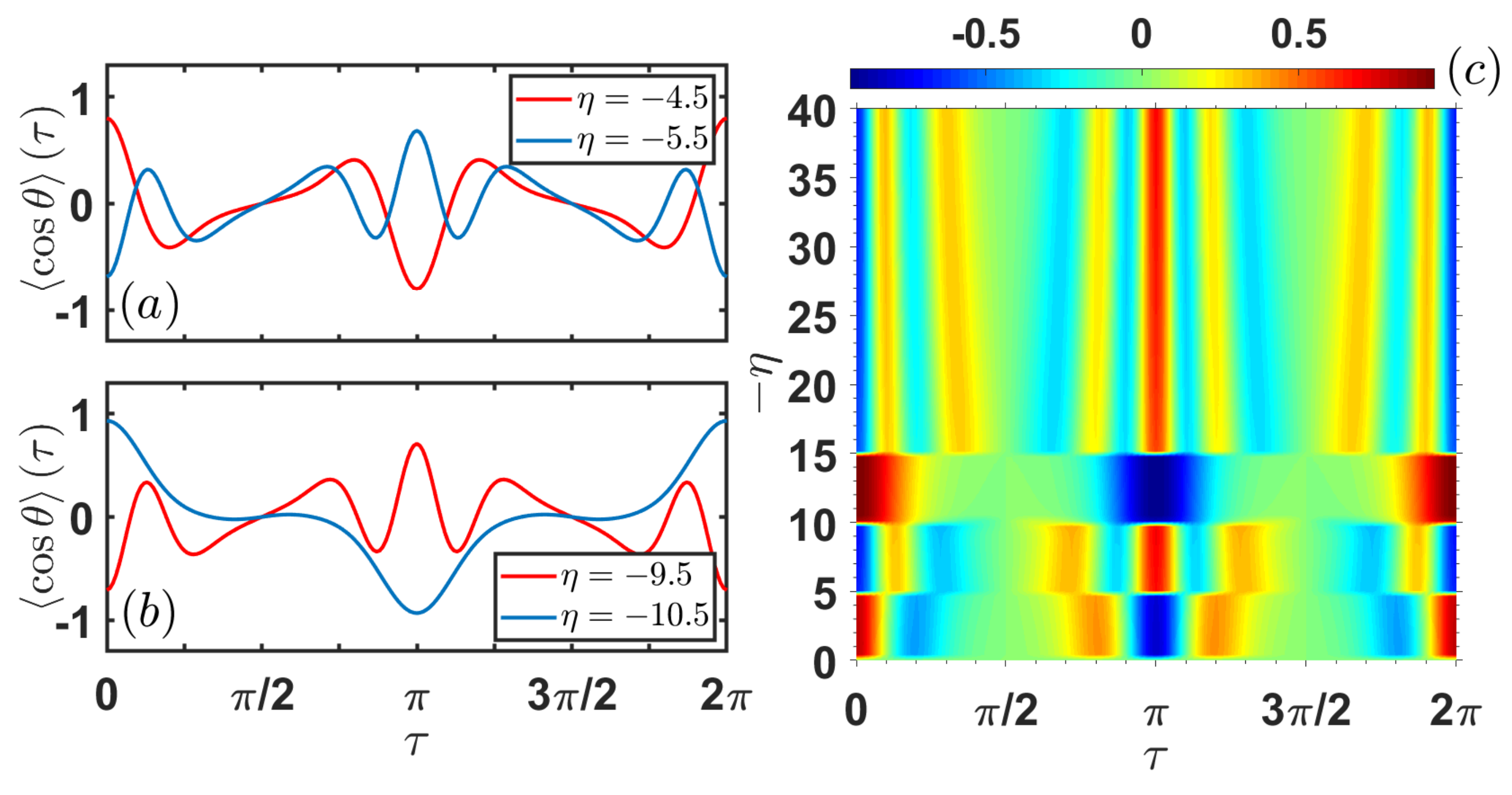}
    	\caption{\label{ori_switchoff}Field-free evolution of the orientation cosine, $\left\langle \cos\theta  \right\rangle(\tau)$, for an initial pendular state $n_0=3$ due to (a) $\zeta = 25$ and $\eta = -4.5,-5.5$; (b) $\zeta=25$ and $\eta= -9.5,-10.5$; and (c) $\zeta=25$ and $\eta\in[-40,0]$. Adapted, with permission, from Ref.~\cite{Mirahmadi2020}.}
    \end{figure}

 A typical post-switch field-free evolution of the orientation, $\left\langle \cos\theta  \right\rangle (\tau)$, is shown in   \cref{ori_switchoff} for the initial pendular state $n_0 = 3$ (the behaviour of other initial pendular states will be addressed at the end of this subsection). 

	Due to the time-dependent phases in \cref{orien_soff}, $\left\langle \cos\theta  \right\rangle (\tau) $ is periodic and found to undergo revivals at the rotational period $\tau_{\mathrm{r}} = 2\pi$ \cite{Mirahmadi2020},
	\begin{align}
		\left\langle \cos\theta  \right\rangle (\tau+2\pi) = \left\langle \cos\theta  \right\rangle (\tau) 
	\end{align} 
    as well as recur with an opposite sign at half of the rotational period, i.e., $\tau = \pi$, 
    \begin{align}
       	\left\langle \cos\theta  \right\rangle (\tau+\pi) = -\left\langle \cos\theta  \right\rangle (\tau) 
    \end{align} 
    Irrespective of the parameters $\eta$ and $\zeta$, the extrema of the orientation cosine (positive or negative) are located at $\tau = \tau_\mathrm{off}$ (defining $\tau=0$ in \cref{ori_switchoff}) and at full and half revival times. Moreover, $\left\langle \cos\theta  \right\rangle$ vanishes at one quarter and three-quarters of $\tau_{\mathrm{r}}$. 
    
        From the time-dependence of \cref{align_soff}, one can show that the alignment recurs after every quarter of the rotational period, 
	\begin{align}
		\left\langle \cos^2\theta  \right\rangle (\tau+\frac{\pi}{2}) = \left\langle \cos^2\theta  \right\rangle (\tau) ~.
	\end{align}

	Let us now discuss the effects of genuine and avoided crossings on the time-evolution of orientation. This is exemplified in panel (c) of \cref{ori_switchoff} which displays the projection onto the $\tau$-$\eta$~plane of $\left\langle \cos\theta  \right\rangle$ for $\zeta=25$ and initial pendular state $n_0=3$. Especially notable is the behaviour at the loci of the avoided and genuine crossings: in the vicinity of the values of the parameters $\eta$ and $\zeta$ that give rise to $\kappa=1,2,3$, the temporal behaviour of the post-switch orientation cosine undergoes a significant variation.  This is shown in more detail in panels (a) and (b) of \cref{ori_switchoff}, which are ``zoom-ins''  of panel (c) near the genuine and avoided crossings located at $\eta = -5$ ($\kappa=1$) and $\eta = -10$ ($\kappa=2$), respectively. 
	 
	The behaviour around the genuine crossings at $\eta = -5$ ($\kappa=1$) and $\eta= -15$ ($\kappa=3$) is a direct consequence of the symmetry swapping, mentioned in \cref{pop_soff},  of the initial pendular states before the turn-off of the combined interaction that alters the coefficients $C_J^{\Gamma,3}$ appearing in \cref{orien_soff}. As shown in \cref{spec}, for either $\kappa<1$ or $\kappa>3$, $\phi_{3}(\theta)$ is of $A_2$ symmetry, whereas for $1<\kappa<3$, $\phi_{3}(\theta)$  is of $A_1$ symmetry.  
	
	A similar trend is observed around the loci of the avoided crossing at $\eta = -10$ ($\kappa=2$). As mentioned previously, although the wavefunction $\phi_{3}(\theta)$ maintains its symmetry ($\Gamma=A_1$ for $1<\kappa<3$), its localization changes dramatically, cf. \cref{avoided}. The concomitant variation of  the $C_J^{\Gamma,3}$ coefficients (around $\kappa=2$) causes the differences in the initial orientation (at $\tau_\mathrm{off}$) as well as in the time-dependence of $\left\langle \cos\theta  \right\rangle$. 
		
	We note that the temporal dependence of alignment closely follows that of orientation. 
	
	 Since the value of $n_0$ gives the number of genuine and avoided crossings of the state $\phi_{n_0}(\theta)$  (see \cref{spec}), it also indicates the number of abrupt changes of the time-dependence of the expectation values, as exemplified for $n_0=3$ in panel (c) of \cref{orien_soff}. 
	 
	 We note that the differences in the temporal behaviour of expectation values corresponding to levels above the maximum of the potential (i.e., of those that approach the free-rotor limit) are much less significant than of those just described, cf. Ref. \cite{Mirahmadi2020}. 
		
	Given that the directional effects discussed above do not occur for either a purely orienting or a purely aligning interaction, we conclude that the combined interaction is \emph{sui generis}, offering additional means to control the post-switch directionality of the system.

	\subsection{Dynamics driven by a rapidly switched-on combined interaction}\label{switch_on}
	We now examine the dynamics of the planar pendulum that arises upon a sudden switch-on of the concurrent orienting and aligning interactions. This case would be brought about by a sudden switch-on of the combined interaction over a time interval $\tau_{\mathrm{on}}-\tau_{\mathrm{off}}\equiv \Delta \tau \ll \tau_{\mathrm{r}}$ followed by its adiabatic switch-off (or its continued presence). Panel (b) of \cref{switches} illustrates this sequence. 
	
	By an argument similar to the one used in the case of the suddenly switched-off interaction, see \cref{switch_off}, $\psi(\theta,\tau_{\mathrm{off}}) = \psi(\theta,\tau_{\mathrm{on}}) $. Given that before the switch-on of the interaction, the system was freely rotating, we choose as the initial state one of the rotational eigenstates, i.e., $\psi(\theta,\tau_{\mathrm{off}}) = \mathrm{e}^{iJ_0\theta}/\sqrt{2\pi}$, with $J_0$ the initial rotational quantum number. The post-switch wavefunction, i.e., for $\tau>\tau_\mathrm{on}$, is then a wavepacket composed of pendular states
	\begin{equation}\label{wf_expan_on}
		\psi(\theta,\tau) =  \sum_{n=0}^{\infty}C^{J_0}_{\Gamma,n}\mathrm{e}^{-i\epsilon_n\tau} \phi_n^{(\Gamma)}(\theta)
	\end{equation} 
	where
\begin{align}\label{C_onn}
		C^{J_0}_{\Gamma,n} = \frac{1}{\sqrt{2\pi}}\int_0^{2\pi}d\theta \mathrm{e}^{iJ_0\theta} \phi_{n}^{(\Gamma)}(\theta)  
	\end{align}
	In \cref{wf_expan_on}, $\Gamma\in\{A_1,A_2\}$ and $\epsilon_n$ and $\phi_n^{(\Gamma)}$ are eigenvalues and eigenfunctions of TISE~\eqref{tise}, i.e, pendular states. Note that the symmetry label $\Gamma$ is absent in $\psi(\theta,\tau)$, as the wavepacket it represents is a mixed superposition of pendular states with symmetries $A_1$ and $A_2$. The expansion coefficients $C^{J_0}_{\Gamma,n}$ are given by \cref{C_A1_son,C_A2_son} in \ref{cal_GJ}, whence it follows that for all initial rotational states $J_0$, the $C^{J_0}_{A_1,n}$ coefficients are real and the $C^{J_0}_{A_2,n}$ coefficients purely imaginary. Moreover, for  $J_0=0$, \cref{C_A2_son} gives $C_{A_2,n}^{0}=0$. 

    We note that even though $C^{J_0}_{A_1,n}$ and $C^{J_0}_{A_2,n}$ are analytically obtainable only for the algebraic pendular states, i.e., for $n<\kappa$, the general forms of the coefficients given by \cref{C_A1_son,C_A2_son} and their properties mentioned above remain in place for any $n$ state, regardless of whether it is analytically obtainable or not.     
      	
	\subsubsection{Population analysis}\label{pop_son}
\begin{figure}
		\centering
		\includegraphics[scale=0.45]{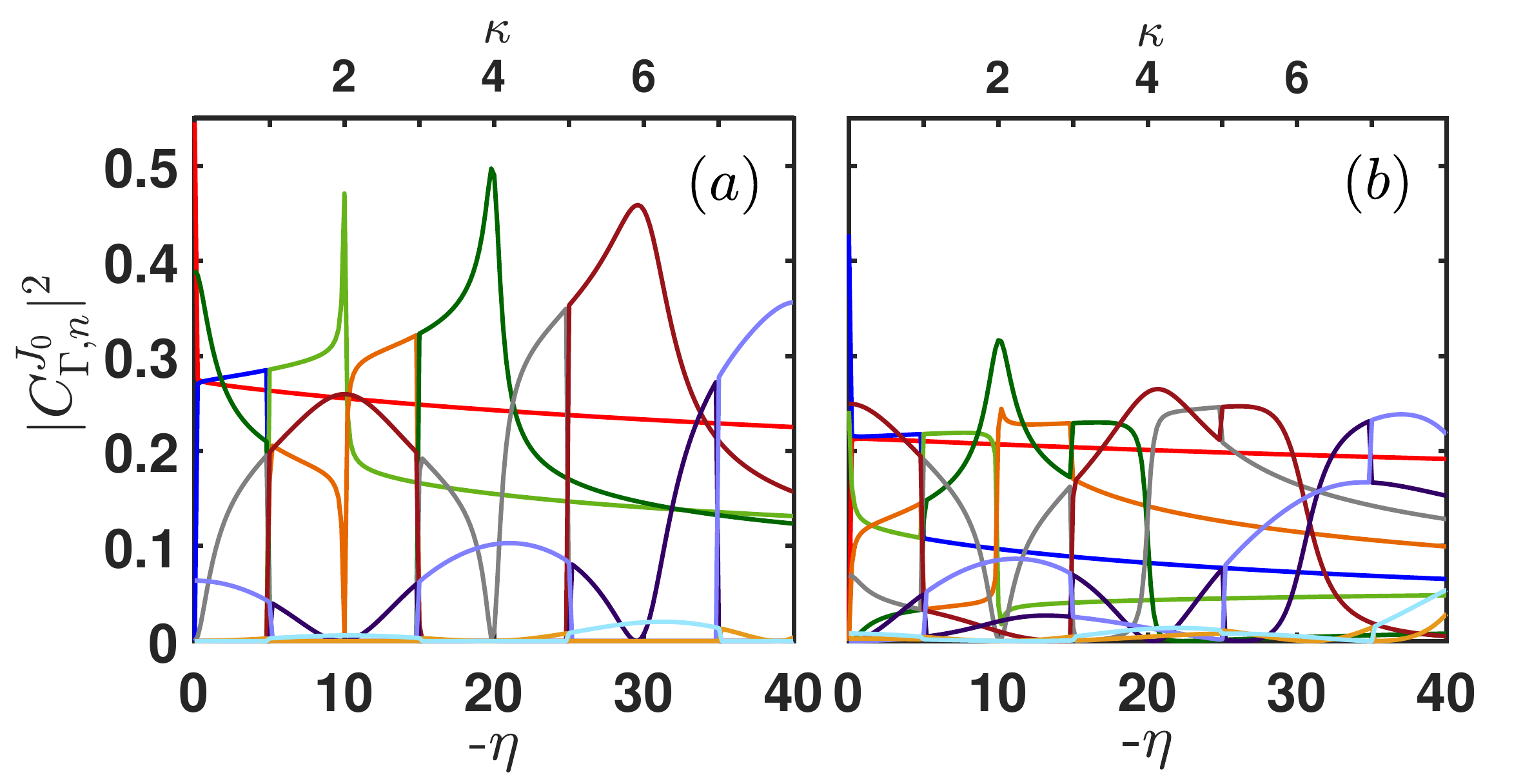}
		\caption{\label{coeff_son_zconst}Populations, $|C^{J_0}_{\Gamma,n}|^2$, of pendular states as functions of $\eta$ at constant $\zeta=25$ after a sudden switch-on of the orienting and aligning interactions. Panels (a) and (b) correspond, respectively, to initial rotational states $J_0 = 0$ and $J_0 = 1$. The upper abscissas show the values of the topological index $\kappa$.  The color-coding of pendular states $n$ is the same as in \cref{spec}. Adopted, with permission, from Ref.~\cite{Mirahmadi2020}.}
	\end{figure}

	We examine two examples of the post-switch populations, $|C^{J_0}_{\Gamma,n}|^2$, as functions of $\eta$ at fixed $\zeta = 25$ that arise from two different initial rotational states, namely $J_0 = 0$ and $J_0 = 1$, see \cref{coeff_son_zconst}.
    The observed rapid variation and sudden flips of the populations at the loci of the avoided and genuine crossings are again related to the overlap integral of the pendular and free-rotor wavefunctions, given here by \cref{C_onn}.     
    
    Let us first consider the population-flips that occur for pairs of $n$ and $n+1$ states at $\eta = -5, -15, -25, -35$, i.e., at genuine crossings arising at odd integer $\kappa$, cf. \cref{coeff_son_zconst}. For $\eta$ and $\zeta$ that give rise to $\kappa < 1$, the pendular states $n = 1$ and $n=2$ pertain to the $\Gamma\equiv A_1$ and $\Gamma \equiv A_2$ symmetry, respectively. However, for $\eta$ and $\zeta$ that correspond to $\kappa > 1$, the states $n=1$ and $n=2$ have symmetries $A_2$ and $A_1$, respectively.  This symmetry swap, $A_1\leftrightarrow A_2$, between the two states results in an interchange of their angular localization and, consequently, of their overlap integrals with the free rotor states and thus of the corresponding populations $|C^{J_0}_{\Gamma,n}|^2$. 
    
    In the vicinity of  $\eta$ and $\zeta$ values that give rise to even integer $\kappa$, i.e., $\eta = -10, -20,-30$ for $\zeta=25$ and hence to avoided crossings, cf. \cref{coeff_son_zconst}, some populations exhibit extrema and others abrupt changes. Although both types of behaviour are a direct consequence of the variation of the pendular eigenfunctions and their overlap with the free rotor wavefunctions, a general explanation applicable to all $n$ and $J_0$ states is elusive. Therefore, in contrast to the behaviour at the genuine crossings, the populations at the avoided crossings have to be treated individually for each state via \cref{C_on}.  
    
    We note that the extrema are more pronounced for $n$ states whose energies are well below the  maximum of the pendular potential and gradually fade away as the pendular eigenvalues come close to the potential's maximum.
    As expected from \cref{C_A2_son}, for $J_0=0$, none of the pendular states of the $\Gamma\equiv A_2$ symmetry are populated, cf. panel (a) of \cref{coeff_son_zconst}. Moreover, for higher initial rotational states (e.g., $J_0>6$ for the parameters considered in \cref{coeff_son_zconst}), only $n$ states whose eigenenergies lie above the potential barrier are populated. Since the wavefunctions $\phi_n$ of these states resemble those of a free rotor,  the sharp variations of the populations are absent for initial rotor states with $J_0\approx n$.

	\subsubsection{Evolution of the expectation values}
	The temporal dependence, $\left\langle S \right\rangle (\tau) \equiv \left\langle \psi(\tau) |S | \psi(\tau) \right\rangle$,  of an observable $S$, such as the directional cosines or the kinetic energy, after rapidly switching on the combined interaction is obtained by substituting from  \cref{wf_expan_on},	
	
	\begin{align}\label{obs_expect}
		\left\langle S \right\rangle (\tau) =& \sum_{n,n'} \left( C^{J_0}_{\Gamma,n}\right)^* C^{J_0}_{\Gamma',n'} e^{i(\epsilon_n-\epsilon_{n'})\tau} \big\langle \phi_n^{(\Gamma)} \big| S \big| \phi_{n'}^{(\Gamma')} \big\rangle \nonumber \\
		=& \sum_{n} |C^{J_0}_{\Gamma,n}|^2 \left\langle \phi_n^{(\Gamma)} \big| S \big| \phi_n^{(\Gamma)} \right\rangle 
		+ 2 \operatorname{Re} \left[  \sum_{n>n'} \left( C^{J_0}_{\Gamma,n} \right)^* C^{J_0}_{\Gamma',n'}  e^{i(\epsilon_n-\epsilon_{n'})\tau} \big\langle \phi_n^{(\Gamma)} \big| S \big| \phi_{n'}^{(\Gamma')}   \big\rangle \right]
	\end{align}  
	with $\Gamma'\in\{A_1,A_2\}$ and $\Gamma\in\{A_1,A_2\}$. 
	By making use of the properties of the coefficients $C^{J_0}_{\Gamma,n}$, we obtain for $S\equiv \cos\theta$	
	\begin{align}\label{orien_son}
		\left\langle \cos\theta\right\rangle(\tau) =&\sum_{n} | C_{\Gamma,n}^{J_0} |^2 \left\langle  \cos\theta \right\rangle_n  + 2  \Bigg[ \sum_{n>n'\in{Q_1}} C_{A_1,n}^{J_0} C_{A_1,n'}^{J_0}  \cos\left[(\epsilon_n-\epsilon_{n'})\tau\right] \left\langle \cos\theta \right\rangle^{(A_1)}_{nn'} \nonumber \\
		& - \sum_{n>n'\in{Q_2}} C_{A_2,n}^{J_0}  C_{A_2,n'}^{J_0}  \cos\left[(\epsilon_n-\epsilon_{n'})\tau\right] \left\langle \cos\theta \right\rangle^{(A_2)}_{nn'} \Bigg]\nonumber \\
		=&  \left\langle \cos\theta \right\rangle_\mathrm{p} + \left\langle \cos\theta \right\rangle_\mathrm{c}(\tau) 
	\end{align}
The selection rules, \cref{srule}, made it possible to break up the summation in the second term of \cref{obs_expect} into two terms, one comprising a sum over the quantum numbers $n\in Q_1$ associated with the $A_1$ symmetry and the other  consisting of $n\in Q_2$ associated with symmetry $A_2$.
The subscripts $\mathrm{p}$ and $\mathrm{c}$ indicate the time-independent ``population'' and time-dependent ``coherence'' terms, respectively. We note that the former terms have $n=n'$ whereas the latter ones have $n \neq n'$; only the ``coherence'' terms are time-dependent. For the sake of simplicity, we did not break the time-independent part  $\left\langle \cos\theta\right\rangle_\mathrm{p}$ into terms with different symmetries. 

We note that \cref{orien_son} includes both numerical and analytic pendular states, although the latter are only available (with substitution from \cref{pen_orien}) if $\kappa$ is an odd integer.    
    
\begin{figure}
    	\centering
    	\includegraphics[scale=0.45]{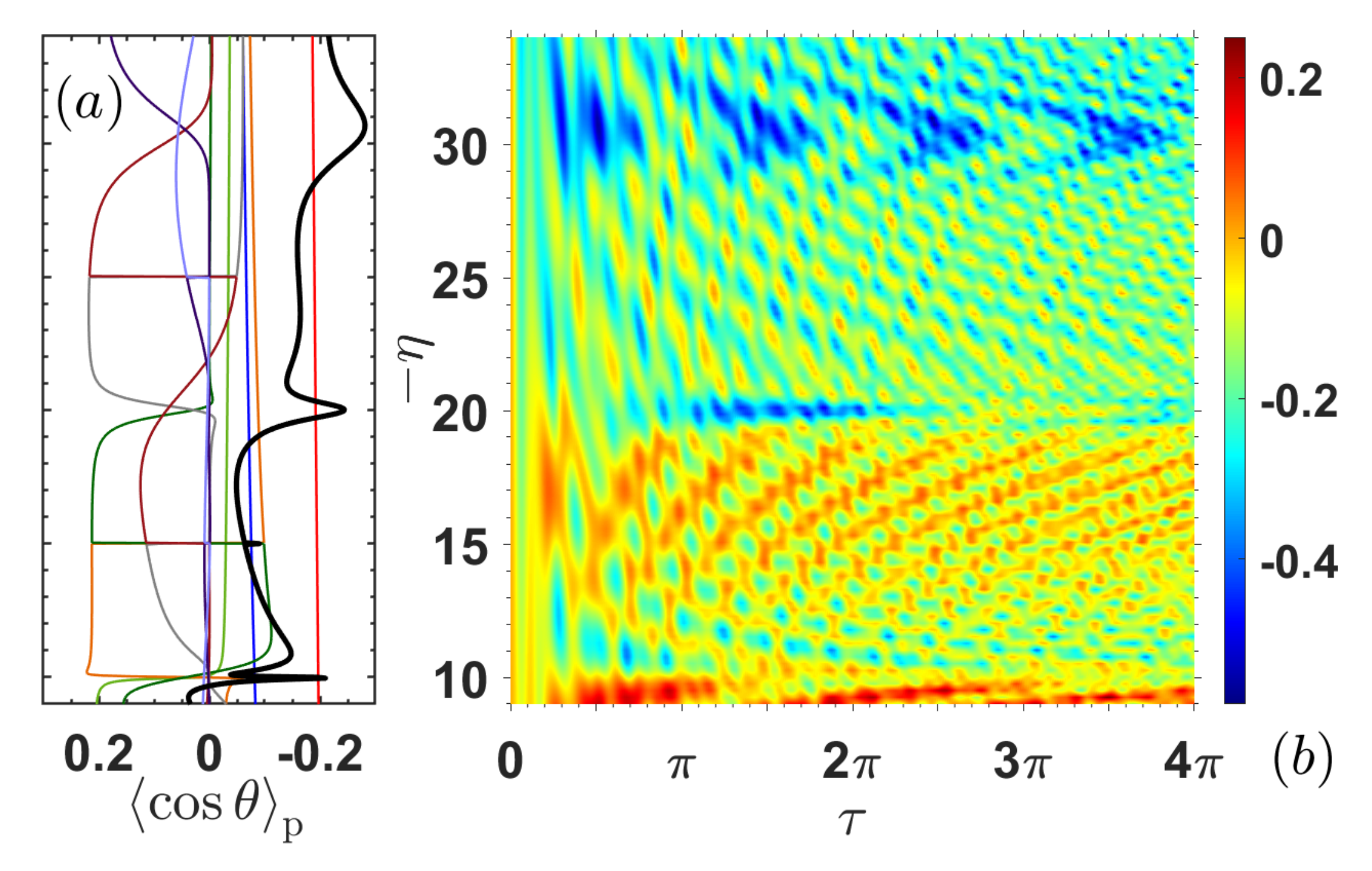} 
    	\caption{ \label{orien_switchon_J1}(a) Time-independent expectation value $\left\langle \cos\theta \right\rangle_\mathrm{p}$ (black curve) for $\zeta=25$ and initial state $J_0=1$, plotted together with the products $\left\langle \cos\theta\right\rangle_n |C_{\Gamma,n}^{J_0=1}|^2$ shown for each state $n$ by the color-coded curves, cf.  \cref{spec}.  (b) Expectation value $\left\langle \cos\theta \right\rangle(\tau)$ for $\zeta=25$,  $-\eta\in[9,34]$, and initial $J_0=1$,  projected onto the $\eta$-$\tau$ plane. The interval $-\eta\in[0,9]$ is omitted for the sake of a better resolution. Adapted, with permission, from Ref.~\cite{Mirahmadi2020}.}
    \end{figure}

The effect of the time-independent term  $\left\langle\cos\theta\right\rangle_\mathrm{p}$ as a function of $\eta$ for a fixed $\zeta=25$ is illustrated in panel (a) of \cref{orien_switchon_J1} for $J_0=1$ (black curve). 
Based on the population analysis of \cref{pop_son} together with the properties of $\left\langle \cos\theta \right\rangle_{n}$ derived from \cref{HF_eta}, shown by the color-coded curves in panel (a) of \cref{orien_switchon_J1}, we would expect significant variation of the products $\left\langle \cos\theta\right\rangle_n |C_{\Gamma,n}^{J_0}|^2$ and, consequently, of $\left\langle \cos\theta\right\rangle_\mathrm{p}$ around the loci of avoided crossings. However, such fluctuations are absent in the vicinity of  genuine crossings where the population as well as the orientation flips between states with different $n$ sum up and cancel each other, see \cref{coeff_son_zconst}. Hence, while $\left\langle \cos\theta\right\rangle_\mathrm{p}$  changes smoothly and rather monotonously in the vicinity of the loci of genuine crossings at $\eta = -15,-25$ in \cref{orien_switchon_J1}, it exhibits significant minima (larger negative orientation) around the loci of avoided crossings; these are apparent at $\eta = -10,-20,-30$. By increasing $|\eta|$ the potential becomes a single-well and the minima of the orientation cosine disappear.  
        
    The coherence term, $\left\langle \cos\theta\right\rangle_\mathrm{c}(\tau)$, in \cref{orien_son} consists of two Fourier sums over terms with frequencies $\Delta\epsilon_{nn'} \equiv \epsilon_{n}-\epsilon_{n'}$. Since $\Delta\epsilon_{nn'}$ is the energy difference between pendular eigenvalues associated with the same symmetry $\Gamma$, genuine crossings do not contribute to the coherence term. As a result, there is no significant dependence of $\left\langle \cos\theta\right\rangle_\mathrm{c}(\tau)$ on the values of the interaction parameters $\eta$ and $\zeta$ that are related by an odd integer $\kappa$. 
    In contrast, at the avoided crossings, i.e., for $\eta$ and $\zeta$ that make $\kappa$ an even integer, the coherence term in \cref{orien_son} comprises frequencies much lower than those that come about at non-integer or odd-integer $\kappa$ values. This results in a longer fundamental period in $\tau$ and, consequently, significantly slower oscillations of $\left\langle \cos\theta\right\rangle(\tau)$ along the loci of avoided crossings.
    
    All these unique features are captured in panel (b) of \cref{orien_switchon_J1}, which shows the projection of $\left\langle \cos\theta \right\rangle(\tau)$ onto the $\eta$-$\tau$ plane for $\zeta=25$,  $-\eta\in[9,34]$, and initial $J_0=1$. The centroids of the fluctuations of the orientation cosine around the loci of the avoided crossings at $\eta = -10,-20,-30$ match accurately the behaviour of $\left\langle \cos\theta\right\rangle_\mathrm{p}(\tau)$ and $\left\langle \cos\theta\right\rangle_\mathrm{c}(\tau)$ outlined above. One can discern the slow but large-amplitude oscillations in $\tau$ along the loci for particular $\eta$ values. For smaller values of $\eta$, the period of oscillations exceeds the rotational period of the rotor, $\tau_\mathrm{r}=2\pi$, and is not captured on a time scale comparable with $\tau_\mathrm{r}$. For instance, for $\eta =-10$, the period is $\approx 32\pi$). 

We note that $\big\langle \cos^2\theta\big\rangle(\tau)$ and $\left\langle \mathbf{J^2}\right\rangle(\tau)$ could be obtained in the same fashion.  For the latter, a simple result follows from energy/angular momentum conservation,  
\begin{align}\label{J0}
\left\langle \mathbf{J^2}\right\rangle(\tau_{\mathrm{on}})=J_0^2
\end{align}
with $J_0$ the rotational quantum number of the initial field-free rotor state.  From the Hamiltonian \eqref{Hdimensionless}, we obtain a closed-form expression for the post-pulse energy,
\begin{align}\label{H_F_1}
\left\langle H(\tau_{\mathrm{on}})\right\rangle=\left\langle \mathbf{J^2}\right\rangle(\tau_{\mathrm{on}})-\eta \left\langle \cos\theta\right\rangle(\tau_{\mathrm{on}})-\zeta \left\langle \cos^2\theta\right\rangle(\tau_{\mathrm{on}})=J_0^2-\frac{1}{2}\zeta
\end{align}
 where we made use  of the identities $\left\langle\cos\theta\right\rangle=0$ and $\left\langle\cos^2\theta\right\rangle=\frac{1}{2}$ that apply to any field-free rotor state $J_0$. Thus upon subjecting the planar rotor to a sudden combined interaction, its energy drops by $\frac{1}{2}\zeta$.
    
\begin{figure}
		\centering
		\includegraphics[scale=0.45]{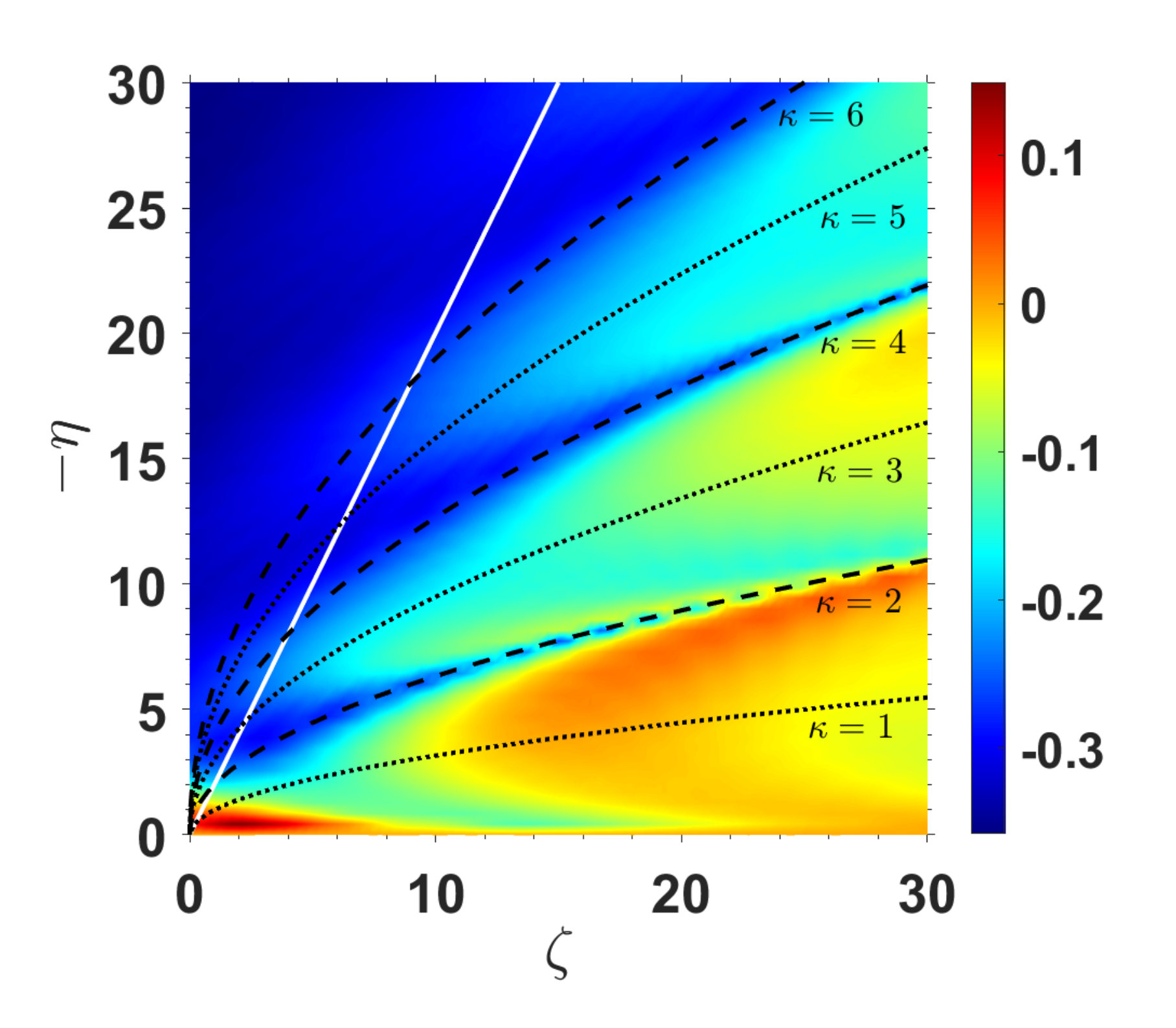} 
		\caption{ \label{ori_ave_4p_son_J1}Projection of the time-averaged orientation cosine, $\overline{\left\langle \cos\theta\right\rangle}_{[0,4\pi]}$ onto the $\eta$-$\zeta$ plane corresponding to a sudden switch-on of the concurrent orienting and aligning interaction. The initial free-rotor state was  $J_0=1$. The black curves are labelled by the values of the topological index $\kappa$, whereby the dotted ones pertain to odd integer $\kappa$ (marking genuine crossings) and the dashed black curves to even integer $\kappa$ (marking avoided crossings). The white line shows the loci of $-\eta = 2\sqrt{\zeta}$, which separates the regimes of single- and double-well potentials. Adapted, with permission, from Ref.~\cite{Mirahmadi2020}.}
	\end{figure}

    Finally, we explore the effects of the crossings on the driven-rotor dynamics by evaluating the time average of the orientation cosine in the entire  $\eta$-$\zeta$ plane, 
 \begin{align}
    	\overline{\left\langle \cos\theta\right\rangle}_{[0,\tilde{\tau}]} =& \int_0^{\tilde{\tau}}\left\langle \cos\theta\right\rangle(\tau)d\tau \nonumber \\
    	=&\tilde{\tau}\left\langle \cos\theta\right\rangle_\mathrm{p} +  2\sum_{n>n'\in{Q_1}} C_{A_1,n}^{J_0} C^{J_0}_{A_1,n'}   \left\langle \cos\theta \right\rangle^{(A_1)}_{nn'} \frac{\sin[(\epsilon_n-\epsilon_{n'})\tilde{\tau}]}{(\epsilon_n-\epsilon_{n'})}\nonumber \\
    	& - 2\sum_{n>n'\in{Q_2}} C_{A_2,n}^{J_0}  C^{J_0}_{A_2,n'}  \left\langle \cos\theta \right\rangle^{(A_2)}_{nn'}\frac{\sin[(\epsilon_n-\epsilon_{n'})\tilde{\tau}]}{(\epsilon_n-\epsilon_{n'})}
    \end{align}   
The time-averaging serves as a smoothing procedure \cite{Mirahmadi2020} and for $\tilde{\tau} = 4\pi$ leads to the projection of the time-averaged orientation cosine shown in \cref{ori_ave_4p_son_J1}. The qualitative behaviour of the time average is found to be independent of $\tilde{\tau}$. 
    
    When applied to the alignment cosine or the kinetic energy, the time-averaging procedure leads to similar results as those shown in \cref{ori_ave_4p_son_J1} (for more details, see Ref.~\cite{Mirahmadi2020}). 
	Thus, as manifested by the time-evolution of the expectation values $\left\langle \cos\theta \right\rangle(\tau)$, $\left\langle \cos^2\theta \right\rangle(\tau)$,  and $\left\langle \mathbf{J}^2\right\rangle(\tau)$, the genuine crossings (dotted curves corresponding to odd values of $\kappa$ in \cref{ori_ave_4p_son_J1}) do not affect the dynamics, whereas avoided crossings (dashed curves corresponding to even values of $\kappa$ in \cref{ori_ave_4p_son_J1}) play havoc with it.  This outcome of our analysis may become the basis for observing the crossings experimentally. 
	
	We note that the behaviour captured by \cref{ori_ave_4p_son_J1} is characteristic of interactions with a double-well potential and fades away at $\eta$-$\zeta$ parameters corresponding to a single-well regime as seen to the left of the white line in \cref{ori_ave_4p_son_J1}.

	\section{Conclusions and prospects}\label{conclusion}
	In our combined analytic and computational study we explored the quantum dynamics of a planar rotor subject to suddenly-switched concurrent orienting and aligning interactions. This study established that observables such as the orientation cosine reflect the topology of the system's eigensurfaces and can, in fact, be used to map them out experimentally. Conversely, as the eigensurface topology affects the dynamics, its knowledge must be incorporated in schemes devised to control the rotor/pendulum system.
	
	The eigensurface topology reflects the system's symmetry properties, which transpire in particular via the genuine or avoided crossings and the angular localization or relocalization of the wavepackets created by the sudden switch-on or switch-off of the combined interaction. Given that the crossings only arise for the combined interaction, the dynamics revealed herein is \emph{sui generis}, with no counterpart in the dynamics due to either the orienting or aligning interaction alone. 
	
Of particular interest are the significant differences in the field-free orientation and alignment of the rotor upon a sudden switch-off of the combined interaction that come about in the vicinity of both genuine and avoided crossings. For the suddenly switched-on interaction, these effects only arise in the vicinity of avoided crossings but are found to cancel out at the genuine crossings. 
 	
Finally, our study paves the way for understanding the full-fledged dynamics of a driven 3D rotor. It is also relevant to the isomorphic problem of manipulating atomic translation in optical traps and lattices, the subject of our ongoing work.
	 
	\section*{Acknowledgements}
	We thank Gerard Meijer for his interest and support and Mallikarjun Karra and Jes\'us P\'{e}rez R\'{i}os for discussions. Support by the Deutsche Forschungsgemeinschaft (DFG) through grants SCHM 1202/3-1 and FR 3319/3-1 as well as DAAD's  STIBET Degree Completion Grant from the Technische Universit\"at  Berlin are gratefully acknowledged.
	
	\appendix
	\section{Evaluation of trigonometric integrals containing $\phi_n^{(\Gamma)}$}\label{pendular_int}
	
	The vanishing transition integrals (selection rules) for orientation and alignment cosines in the basis of pendular states follow from \cref{eig_A}: 
	\begin{align}\label{srule}
		\big\langle \phi_n^{(A_1)} \big| \cos\theta \big| \phi_{n'}^{(A_2)} \big\rangle &= \big\langle \phi_n^{(A_2)} \big| \cos\theta \big| \phi_{n'}^{(A_1)} \big\rangle  = 0 ~, \nonumber \\
		\big\langle \phi_n^{(A_1)} \big| \cos^2\theta \big| \phi_{n'}^{(A_2)} \big\rangle &= \big\langle \phi_n^{(A_2)} \big| \cos^2\theta \big| \phi_{n'}^{(A_1)} \big\rangle  = 0 
	\end{align}
	The non-vanishing transition integrals $\langle\cos\theta \rangle_{nn'}^{(\Gamma)} \equiv\big\langle \phi_n^{(\Gamma)} \big| \cos\theta \big| \phi_{n'}^{(\Gamma)} \big\rangle $ and  $\langle\cos^2\theta \rangle_{nn'}^{(\Gamma)} \equiv\big\langle \phi_n^{(\Gamma)} \big| \cos^2\theta \big| \phi_{n'}^{(\Gamma)} \big\rangle $ are given below.
    	
	In order to obtain the expectation values of the orientation and alignment cosines, we need to evaluate the integrals of the following form (note that $\phi_n^{(\Gamma)}$s are real functions)
	\begin{equation}\label{f_int}
		\int_{0}^{2\pi} d\theta e^{2\sqrt{\zeta}\cos\theta} f(\theta) \cos^{2L}\frac{\theta}{2} 
	\end{equation}
    with $L\coloneqq\ell+\ell'$ and $f(\theta)$ being one of the functions $\cos\theta$ or $\cos^2\theta$. As an example, we calculate integral~\eqref{f_int} for the case of $f(\theta) = \cos\theta$. By making use of the trigonometric relation $\cos^{2L}(\theta/2)=1/(2^{2L})\left(\binom{2L}{L} + \sum_{m=0}^{L-1} 2\binom{2L}{m}\cos[(L-m)\theta] \right)$ \cite{Gradshteyn2007} and the product relation of two cosine functions, we obtain     
	\begin{align}\label{integralform}
		\int_{0}^{2\pi} d\theta e^{2\sqrt{\zeta}\cos\theta} \cos\theta \cos^{2L}\frac{\theta}{2} =& \frac{1}{2^{2L}} \Bigg\{ \binom{2L}{L}\int_{0}^{2\pi} d\theta e^{2\sqrt{\zeta}\cos\theta} \cos\theta \nonumber \\
		& + 2  \sum_{m=0}^{L-1} \binom{2L}{m} \int_{0}^{2\pi} d\theta e^{2\sqrt{\zeta}\cos\theta} \cos\theta \cos[(L-m)\theta] \Bigg\} \nonumber \\
		= & \frac{2\pi}{2^{2L}}  \Bigg\{ \binom{2L}{L} I_1(2\sqrt{\zeta})  \nonumber \\
		&+  \sum_{m=0}^{L-1} \binom{2L}{m} \left[I_{L-m-1}(2\sqrt{\zeta}) + I_{L-m+1}(2\sqrt{\zeta}) \right]\Bigg\} 
	\end{align}
	where $I_{\rho}$ is the modified Bessel function of the first kind of the $\rho$-th order (for integer $\rho$) and $\binom{b}{a}$ is the binomial coefficient. Hence we have \cite{Mirahmadi2020}
	\begin{align}\label{pen_orien}
		\left\langle  \cos\theta  \right\rangle^{(A_1)}_{nn'} =& \sum_{\ell,\ell'}\frac{g_{nn'\ell\ell'}^{(A_1)}  }{2^{2L}}  \Bigg( \binom{2L}{L} I_1(2\sqrt{\zeta}) 
		+  \sum_{m=0}^{L-1} \binom{2L}{m} \left[I_{L-m-1}(2\sqrt{\zeta}) + I_{L-m+1}(2\sqrt{\zeta}) \right]  \Bigg) \nonumber \\
		\left\langle  \cos\theta  \right\rangle^{(A_2)}_{nn'}   =& \sum_{\ell,\ell'}\frac{g_{nn'\ell\ell'}^{(A_2)}  }{2^{2L+2}} \frac{1}{\sqrt{\zeta}}\Bigg( \binom{2L}{L} 2I_2(2\sqrt{\zeta})  + \sum_{m=0}^{L-1} \binom{2L}{m} \Big[ (L-m+2)I_{L-m+2}(2\sqrt{\zeta}) \nonumber \\
		& - (L-m-2)I_{L-m-2}(2\sqrt{\zeta})\Big]  \Bigg) 
	\end{align}
	where $g_{nn'\ell\ell'}^{(A_1)}=2\pi v_{n,\ell} v_{n',\ell'}/\sqrt{N_n^{(A_1)}N_{n'}^{(A_1)}}$ and $g_{nn'\ell\ell'}^{(A_2)}=2\pi \tilde{v}_{n,\ell} \tilde{v}_{n',\ell'}/\sqrt{N_n^{(A_2)}N_{n'}^{(A_2)}}$. Similarly, for the alignment cosine ($f(\theta)=\cos^2\theta$) we can write 
	\begin{align}\label{pen_align}
		\left\langle  \cos^2\theta  \right\rangle^{(A_1)}_{nn'}  =& \sum_{\ell,\ell'} \frac{g_{nn'\ell\ell'}^{(A_1)} }{2^{2L+1}} \Bigg( \binom{2L}{L}\left[ I_0(2\sqrt{\zeta}) + I_2 (2\sqrt{\zeta}) \right] + \sum_{m=0}^{L-1} \binom{2L}{m} \Big[2 I_{L-m}(2\sqrt{\zeta})  \nonumber \\
		& + I_{L-m+2}(2\sqrt{\zeta}) + I_{L-m-2}(2\sqrt{\zeta}) \Big] \Bigg) ~, \nonumber \\
		\left\langle  \cos^2\theta  \right\rangle^{(A_2)}_{nn'}  =& \sum_{\ell,\ell'} \frac{g_{nn'\ell\ell'}^{(A_2)} }{2^{2L+3}}  \Bigg(  \binom{2L}{L}\left[ I_0(2\sqrt{\zeta}) - I_4 (2\sqrt{\zeta}) \right] + \sum_{m=0}^{L-1} \binom{2L}{m} \Big[2 I_{L-m}(2\sqrt{\zeta}) \nonumber \\
		&  + I_{L-m+4}(2\sqrt{\zeta}) - I_{L-m-4}(2\sqrt{\zeta}) \Big] \Bigg) 
	\end{align}
	
	In a similar manner, the normalization factors $N^{(A_1)}_{n}$ and $N^{(A_2)}_{n}$ in \cref{eig_A} are obtained by substituting $1$ and $\sin\theta$ for $f(\theta)$ in Eq. \eqref{f_int}, respectively, as 
	\begin{align}\label{N_A1A2}
		N^{(A_1)}_{n} = & 2\pi\sum_{\ell,\ell'} \frac{1 }{2^{2L}} v_{n,\ell} v_{n,\ell'} \left(   \binom{2L}{L} I_0(2\sqrt{\zeta}) + 2\sum_{m=0}^{L-1} \binom{2L}{m} I_{L-m}(2\sqrt{\zeta}) \right) ~, \nonumber \\
		N^{(A_2)}_{n} = & 2\pi\sum_{\ell,\ell'} \frac{1 }{2^{2L+1}} \tilde{v}_{n,\ell} \tilde{v}_{n,\ell'} \Bigg(   \binom{2L}{L} I_1(2\sqrt{\zeta})/\sqrt{\zeta} + \sum_{m=0}^{L-1} \binom{2L}{m} \left[ 2I_{L-m}(2\sqrt{\zeta})  \right.  \nonumber \\
		&  \left. - I_{L-m-2}(2\sqrt{\zeta}) - I_{L-m+2}(2\sqrt{\zeta})\right]  \Bigg) 
	\end{align} 

	\section{Evaluation of the expansion coefficients in the post-switch wavepackets}\label{cal_GJ}
	\subsection*{The case of rapidly switched-off fields}
	The coefficients $C^{\Gamma,n_0}_J$ in \cref{wf_expan_off} can be obtained through the standard expansion theorem as \cite{Mirahmadi2020}
	\begin{align}\label{C_off}
		C^{\Gamma,n_0}_J = \frac{1}{\sqrt{2\pi}}\int_0^{2\pi}d\theta \mathrm{e}^{-iJ\theta} \phi_{n_0}^{(\Gamma)}(\theta)  
	\end{align}
	Upon substituting for $\phi_{n0}^{(A_1)}$ from \cref{eig_A} into \cref{C_off},  the coefficients for $\Gamma \equiv A_1$ can be calculated via integrals of the form
	\begin{align}\label{int_A1_2}
		\int_0^{2\pi}d\theta \mathrm{e}^{\sqrt{\zeta}\cos\theta} \cos(J\theta) \cos^{2\ell}\frac{\theta}{2} =& \int_0^{2\pi}d\theta\mathrm{e}^{\sqrt{\zeta}\cos\theta}  \cos(J\theta) \cos^{2\ell}\frac{\theta}{2} \nonumber \\
		= & \frac{2\pi}{2^{2\ell}} \Bigg[\sum_{m=0}^{\ell-1} \binom{2\ell}{m} \left( I_{J+\ell-m}(\sqrt{\zeta}) + I_{J-\ell+m}(\sqrt{\zeta}) \right)  \nonumber \\ 
		& + \binom{2\ell}{\ell} I_{J}(\sqrt{\zeta}) \Bigg] 
	\end{align}
	where the imaginary part of the integral, $i\sin(J\theta)$, vanishes due to symmetry and where we again made use of the trigonometric relation  $\cos^{2\ell}(\theta/2)=1/(2^{2\ell})\left(\binom{2\ell}{\ell} + \sum_{m=0}^{\ell-1} 2\binom{2\ell}{m}\cos[(\ell-m)\theta] \right)$ and the definition of $I_{\rho}$, i.e., the modified Bessel function of the first kind and $\rho$-th order \cite{Gradshteyn2007}. 	
	Therefore, the coefficients $C^{A_1,n_0}_J $ in \cref{wf_expan_off} can be written as follows
	\begin{align}\label{C_A1}
		C^{A_1,n_0}_J = & \sqrt{\frac{2\pi}{N^{(A_1)}_{n_0}}}\sum_{\ell=0}^{\infty} \frac{v_{n_0,\ell}}{2^{2\ell}} \Bigg[  \sum_{m=0}^{\ell-1} \binom{2\ell}{m} \left( I_{J+\ell-m}(\sqrt{\zeta}) + I_{J-\ell+m}(\sqrt{\zeta}) \right)  \nonumber \\
		& + \binom{2\ell}{\ell} I_J(\sqrt{\zeta}) \Bigg] 
	\end{align} 
	
	The coefficients associated with a pendular state pertaining to the $A_2$ symmetry obtains from integrals of the form
	\begin{align}\label{int_A2}
		\int_0^{2\pi}d\theta \mathrm{e}^{-iJ\theta} \mathrm{e}^{\sqrt{\zeta}\cos\theta} \sin\theta\cos^{2\ell}\frac{\theta}{2} =
		-i\int_0^{2\pi}d\theta \mathrm{e}^{\sqrt{\zeta}\cos\theta} \sin(J\theta) \sin\theta\cos^{2\ell}\frac{\theta}{2}   
	\end{align} 
	where the real part of the integral vanishes by symmetry. Using the relation for the product of two sine functions and following a procedure similar to that used in deriving \cref{int_A1_2} we obtain
	\begin{align}\label{C_A2}
		C^{A_2,n_0}_J = &i \sqrt{\frac{2\pi}{N^{(A_2)}_{n_0}}} \sum_{\ell=0}^{\infty} \frac{\tilde{v}_{n_0,\ell}}{2^{2\ell+1}} \Bigg[  \sum_{m=0}^{\ell-1} \binom{2\ell}{m} \Big( I_{J+1+\ell-m}(\sqrt{\zeta}) + I_{J+1-\ell+m}(\sqrt{\zeta})  \nonumber \\
		&- I_{J-1+\ell-m}(\sqrt{\zeta}) - I_{J-1-\ell+m}(\sqrt{\zeta}) \Big)  + \binom{2\ell}{\ell} \left( I_{J+1} (\sqrt{\zeta}) - I_{J-1} (\sqrt{\zeta})\right) \Bigg] 
	\end{align} 
	From the properties of the modified Bessel function in \cref{C_A1,C_A2}, we obtain $C^{A_1,n_0}_{-J} = C^{A_1,n_0}_{J} $ and $C^{A_2,n_0}_{-J} = -C^{A_2,n_0}_{J} $. Thus $C^{A_2,n_0}_0$ is always zero, whereas $C^{A_1,n_0}_{0}$ is not \cite{Mirahmadi2020}. 	
	Another key difference between the coefficients associated with symmetries $A_1$ and $A_2$ is that $C^{A_1,n_0}_{J}$ are real and $C^{A_2,n_0}_{J}$ are purely imaginary.

	\subsection*{The case of rapidly switched-on fields}
	By making use of orthonormality of pendular states, it is possible to derive the coefficients $C^{J_0}_{\Gamma,n}$ in \cref{wf_expan_on} via integral (note that $\phi_{n}=\phi^*_n$)
	\begin{align}\label{C_on}
		C^{J_0}_{\Gamma,n} = \frac{1}{\sqrt{2\pi}}\int_0^{2\pi}d\theta \mathrm{e}^{iJ_0\theta} \phi_{n}^{(\Gamma)}(\theta)  
	\end{align}
	Like for the rapidly switched-off fields, we find 
	\begin{align}\label{C_A1_son}
		C^{J_0}_{A_1,n} = & \sqrt{\frac{2\pi}{N^{(A_1)}_{n}}}\sum_{\ell=0}^{\infty} \frac{v_{n,\ell}}{2^{2\ell}} \Bigg[  \sum_{m=0}^{\ell-1} \binom{2\ell}{m} \left( I_{J_0+\ell-m}(\sqrt{\zeta}) + I_{J_0-\ell+m}(\sqrt{\zeta})  \right) \nonumber \\
		& + \binom{2\ell}{\ell} I_{J_0}(\sqrt{\zeta})  \Bigg] 
	\end{align} 
    and 
	\begin{align}\label{C_A2_son}
		C^{J_0}_{A_2,n} =& i\sqrt{\frac{2\pi}{N^{(A_2)}_{n}}}\sum_{\ell=0}^{\infty} \frac{\tilde{v}_{n,\ell}}{2^{2\ell+1}}  \Bigg[  \sum_{m=0}^{\ell-1} \binom{2\ell}{m} \Big( I_{J_0-1+\ell-m}(\sqrt{\zeta}) + I_{J_0-1-\ell+m}(\sqrt{\zeta})    \nonumber \\
		& - I_{J_0+1+\ell-m}(\sqrt{\zeta}) - I_{J_0+1-\ell+m}(\sqrt{\zeta}) \Big) + \binom{2\ell}{\ell} \Big( I_{J_0-1} (\sqrt{\zeta}) - I_{J_0+1} (\sqrt{\zeta})\Big) \Bigg] 
	\end{align}
	for the pendular states with $\Gamma\equiv A_1$ and $\Gamma\equiv A_2$, respectively.
	
	\section{Variation of the expectation values via the Hellmann-Feynman theorem}\label{H_F}
	Of particular importance is the behaviour of the expectation values $\langle\cos\theta \rangle_{n} \equiv\big\langle \phi_n^{(\Gamma)} \big| \cos\theta \big| \phi_{n}^{(\Gamma)} \big\rangle $ and $\langle\cos^2\theta \rangle_{n} \equiv\big\langle \phi_n^{(\Gamma)} \big| \cos^2\theta \big| \phi_{n}^{(\Gamma)} \big\rangle $ near the loci of the  crossings. In tackling the variations of the directional cosines, we make use of the Hellmann-Feynman theorem, $\left\langle \phi_{n} \left| \partial_\lambda H(\lambda) \right| \phi_{n} \right\rangle = \partial_\lambda \epsilon_{n}$,  whence we obtain \cite{Mirahmadi2020},
	\begin{align}\label{HF_eta}
		\left\langle  \cos\theta  \right\rangle_n = - \frac{\partial \epsilon_n}{\partial\eta} ~, \quad
		\left\langle  \cos^2\theta  \right\rangle_n = - \frac{\partial \epsilon_n}{\partial\zeta}  
	\end{align}
    Thus the Hellmann-Feynman theorem, \cref{HF_eta}, suggests that significant variation of the orientation and alignment cosines will take place at the loci of the crossings, see Ref.~\cite{Mirahmadi2020} for further details. 
    
Similarly significant  changes of the expectation value of the kinetic energy $\left\langle \mathbf{J}^2 \right\rangle_{n}$ (in units of the rotational constant $B$) are to be expected at the crossings, given that for a fixed pair of $\eta$ and $\zeta$ and a quantum state $\phi_{n}^{(\Gamma)}$,  
    \begin{equation}\label{kin_bound}
    	\left\langle \mathbf{J}^2 \right\rangle_{n} = \epsilon_n+ \eta \left\langle  \cos\theta  \right\rangle_n + \zeta \left\langle  \cos^2\theta  \right\rangle_n 
    \end{equation}

	\bibliographystyle{iopart-num-href}
	\bibliography{Planar_Rotor_Switched.bib}

\end{document}